\documentclass[runningheads]{llncs}
\usepackage[T1]{fontenc}
\usepackage{cite}
\usepackage{amsmath,amssymb,amsfonts}
\usepackage{algorithmic}
\usepackage{graphicx}
\usepackage{textcomp}
\usepackage{xcolor}
\usepackage{url}
\usepackage{paralist}

\usepackage{booktabs}       
\usepackage{array}          
\usepackage[table]{xcolor}   
\usepackage{ragged2e} 
\usepackage[most]{tcolorbox}

\usepackage{tikz}
\usetikzlibrary{positioning, shapes.geometric, arrows.meta, fit}

\begin{document}
\title{Prompts as Software Engineering Artifacts: \\A Research Agenda and Preliminary Findings}
\titlerunning{Prompts as Software Engineering Artifacts}
%
\author{Hugo Villamizar\inst{1}\orcidID{0000-0003-4142-6967} \and
Jannik Fischbach\inst{2,1}\orcidID{0000-0002-4361-6118} \and
Alexander Korn\inst{3}\orcidID{0009-0002-6258-6791} \and
Andreas Vogelsang\inst{3}\orcidID{0000-0003-1041-0815} \and
Daniel Mendez\inst{4,1}\orcidID{0000-0003-0619-6027}
}

\authorrunning{H. Villamizar et al.}

\institute{
fortiss GmbH, Munich, Germany \\ \email{guarinvillamizar@fortiss.org} \and
Netlight Consulting, Munich, Germany \\ \email{jannik.fischbach@netlight.com} \and
University of Duisburg-Essen, Essen, Germany \\ \email{\{alexander.korn, andreas.vogelsang\}@uni-due.de} \and
Blekinge Institute of Technology, Karlskrona, Sweden \\ \email{daniel.mendez@bth.se}
}

\maketitle              
\begin{abstract} 
Developers now routinely interact with large language models (LLMs) to support a range of software engineering (SE) tasks. This prominent role positions prompts as potential SE artifacts that, like other artifacts, may require systematic development, documentation, and maintenance. However, little is known about how prompts are actually used and managed in LLM-integrated workflows, what challenges practitioners face, and whether the benefits of systematic prompt management outweigh the associated effort. To address this gap, we propose a research programme that (a) characterizes current prompt practices, challenges, and influencing factors in SE; (b) analyzes prompts as software artifacts, examining their evolution, traceability, reuse, and the trade-offs of systematic management; and (c) develops and empirically evaluates evidence-based guidelines for managing prompts in LLM-integrated workflows. As a first step, we conducted an exploratory survey with 74 software professionals from six countries to investigate current prompt practices and challenges. The findings reveal that prompt usage in SE is largely ad-hoc: prompts are often refined through trial-and-error, rarely reused, and shaped more by individual heuristics than standardized practices. These insights not only highlight the need for more systematic approaches to prompt management but also provide the empirical foundation for the subsequent stages of our research programme.

\keywords{software engineering \and AI engineering \and prompt management}
\end{abstract}

\section{Introduction}
\label{sec:introduction}

Large language models (LLMs) are now integral to software engineering (SE) tasks, supporting code generation, debugging, testing, and documentation~\cite{stackoverflow2024survey}. As prompts serve as the sole mechanism for interfacing with LLMs, their role and importance in SE workflows have expanded, raising the question of how they should be positioned within these workflows.

In SE, an artifact is a self‑contained work result with a context‑specific purpose, comprising a physical representation, syntactic structure, and semantic content~\cite{mendez2019artefacts}. From this definition, prompts arguably fulfill the role of artifacts: they encapsulate developer intent, embody critical design decisions, and evolve iteratively as developers refine them to improve LLM outputs. Moreover, they appear in varied contexts, from embedded in source code to detached environments (\textit{e.g.}, ChatGPT) or context‑aware tools (\textit{e.g.}, Copilot).

Over the past few years, research on prompts has gained attention across multiple disciplines. Studies have explored strategies for crafting effective prompts~\cite{white2024chatgpt, liu2022design}, developed interactive tools to support their refinement and evaluation~\cite{kim2024evallm, mishra2023promptaid, cheng2024prompt}, examined how non-expert users approach the task~\cite{zamfirescu2023johnny}, and exploring meta-cognitive demands in LLMs usage~\cite{tankelevitch2024metacognitive}. In SE, prompts has been investigated in contexts such as traceability, testing, and pattern use~\cite{ronanki2024requirements, rodriguez2023prompts, vogelsang2025impact}, as well as in terms of its evolution and quality concerns~\cite{tafreshipour2025prompting, chen2025promptware}. Some work has even framed prompts as “programs,” developing an understanding of prompt programming~\cite{liang2025prompts}.

However, existing research largely addresses isolated aspects without offering a holistic, empirically grounded understanding of how prompts should be systematically managed across LLM‑integrated workflows. In particular, little is known about their long‑term evolution, traceability, and reuse, as well as the trade‑offs between the effort of managing them and the benefits this brings. We also lack empirical insight into how software professionals perceive the process of treating prompts as first‑class SE artifacts.

To advance a systematic understanding of prompts as SE artifacts, this work makes two main contributions:

\begin{compactitem}
    \item \textbf{A research agenda} for investigating prompts as SE artifacts, focusing on three key aspects: (i) characterizing current prompt practices, challenges, and influencing factors; (ii) analyzing prompts in terms of their evolution, traceability, and reuse; and (iii) developing and empirically evaluating guidelines for their systematic management.
    
    \item \textbf{Emerging empirical insights from an exploratory survey} with 74 software professionals, offering early evidence on practices and challenges relevant to managing prompts as SE artifacts in LLM‑integrated workflows.

\end{compactitem}

In this work, we distinguish two complementary dimensions of working with prompts. \textit{Prompt engineering} is the technical process of designing and refining prompts to achieve desired LLM outputs, covering early stages such as specifying intent, structuring text, and embedding prompts in workflows. \textit{Prompt management}, in contrast, is the broader set of practices to keep prompts effective, maintainable, and reusable over time, including versioning, and tracing links to other artifacts (\textit{e.g.}, requirements, code). Together, they form a \textit{prompt life‑cycle} from initial specification to long‑term reuse. While both matter, this work focuses on the management dimension and how they can be systematically used in LLM‑integrated workflows.
\section{Research Agenda}
\label{sec:research-agenda}

\begin{figure}
    \centering
    \includegraphics[width=1\textwidth]{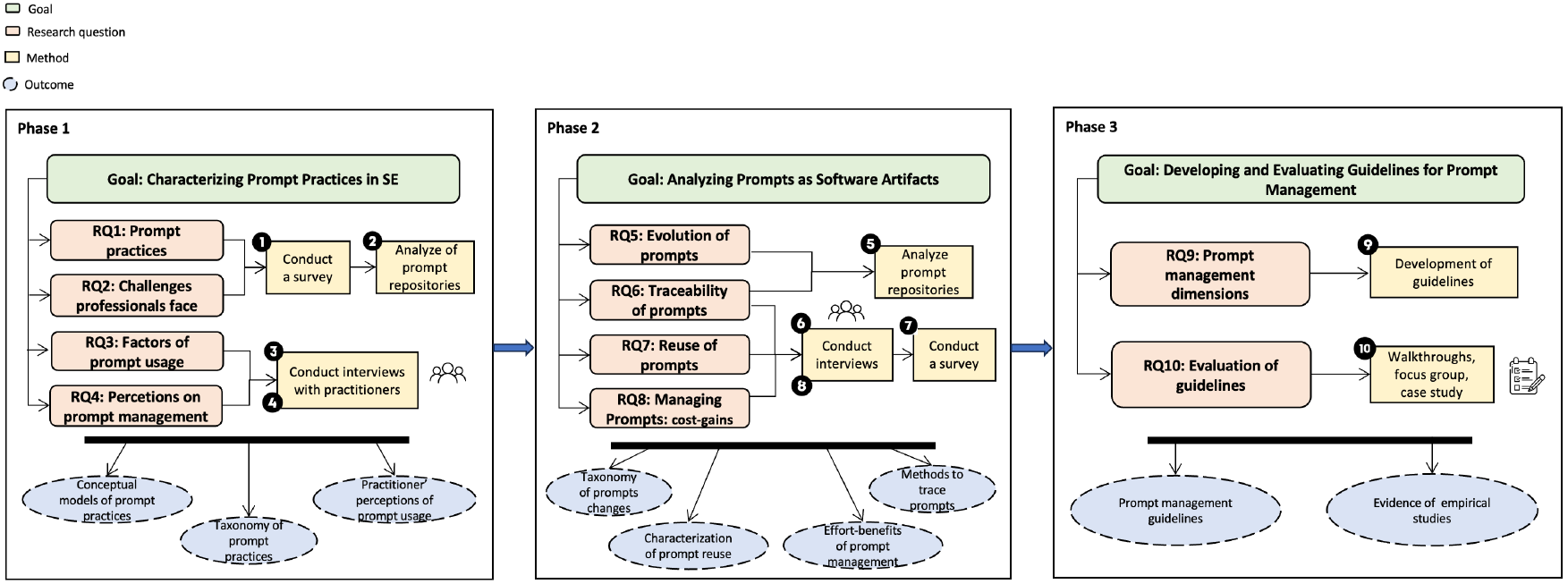} 
    \caption{Overview of our proposed research agenda.}
    \label{fig:research-agenda}
\end{figure}

Our long-term objective is to provide a systematic empirical foundation for understanding the role of prompts as SE artifacts and establish evidence-based practices for managing them systematically in LLM-integrated workflows. Figure~\ref{fig:research-agenda} illustrates the overall structure of the research agenda.

\subsection{Characterizing Prompt Practices in SE}

Phase~1 of our research agenda focuses on understanding how professionals currently interact with prompts in the context of SE tasks and what challenges and factors shape these practices. The following research questions (RQs) guide this phase:

\begin{compactitem}
    \item \textbf{RQ1} What prompt practices are currently followed by software professionals in SE contexts?

    \item \textbf{RQ2} What challenges do software professionals face when using and managing prompts in SE contexts?

    \item \textbf{RQ3} What factors influence the way prompts are created, adapted, and reused in SE workflows?

    \item \textbf{RQ4} How do software professionals perceive the process of managing prompts as software artifacts? 
\end{compactitem}

To address these questions, we first conducted an exploratory survey, which enabled us to identify current prompt practices and challenges encountered in SE contexts (\textit{Cf.} Section~\ref{sec:results}). Second, we plan to analyze publicly available prompt repositories and datasets such as PromptBase~\cite{promptbase}, PromptSet~\cite{pister2024promptset} and DevGPT \cite{xiao2024devgpt} to identify recurring patterns and strategies. Finally, we will conduct semi-structured
interviews with practitioners to (i) understand what drives their prompt choices (\textit{e.g.}, task complexity, prior experience with LLMs, team norm) and (ii) whether they see value in treating prompts more systematically. 


\subsection{Analyzing Prompts as SE Artifacts}

Phase~2 builds on the results of Phase~1 to investigate the extent to which prompts exhibit characteristics of SE artifacts and what this implies for their systematic management. We will focus on understanding prompt evolution, traceability, reuse, and the effort‑benefit trade‑off involved in managing prompts in a structured way. The following RQs guide this phase:

\begin{compactitem}
    \item \textbf{RQ5} How do prompts change throughout a software project, and what patterns of evolution can be observed?
    
    \item \textbf{RQ6} How can we establish traceability between prompts, the LLM‑generated artifacts they produce, and downstream software components?
    
    \item \textbf{RQ7} What mechanisms support effective storage, indexing, and reuse of prompts across projects and teams in SE?
    
    \item \textbf{RQ8} What are the trade‑offs between the effort of managing prompts and the benefits in terms of maintainability, reuse, and confidence in LLM‑assisted development?
\end{compactitem}

To address these questions, we will first conduct repository mining to analyze prompt evolution and traceability, examining changes across commits and their links to input and output artifacts. Second, we will perform interviews with practitioners to capture their mental models, strategies, and pain points around prompt change, reuse, and traceability. Third, we will complement these findings with a survey to quantify reuse and storage practices across different SE settings. Finally, we will conduct targeted interviews or focus groups to explore how professionals perceive the effort‑benefit trade‑off of systematic prompt management.


\subsection{Developing and Evaluating Guidelines for Prompt Management}

Phase~3 builds on the empirical insights from Phases~1 and~2 to synthesize a set of evidence‑based guidelines for the systematic management of prompts in LLM‑integrated workflows. The following RQs guide this phase:

\begin{compactitem}
    \item \textbf{RQ9} What practices and recommendations support the systematic management of prompts in LLM-integrated workflows?
    
    \item \textbf{RQ10} How do software professionals perceive the usefulness and applicability of these prompt management guidelines in practice?
\end{compactitem}

To address these questions, we will synthesize findings from Phases~1 and~2 to distill recurring challenges, effective practices, and practitioner needs into a structured set of guidelines. The synthesis will follow established practices for creating empirical SE guidelines, with each recommendation specifying its application context, the core advice, and the rationale supported by evidence. Guidelines will be enriched with illustrative examples from our data (\textit{e.g.}, documenting prompt intent, maintaining version history, modularizing prompts for reuse). The guidelines will then be evaluated through a combination of structured walkthroughs, focus groups, and field‑based case studies.

\section{Exploratory Survey Design}

We conducted an exploratory, questionnaire-based online survey, following established guidelines for survey research in SE~\cite{ciolkowski2003practical}. The aim was to obtain early insights into how software professionals use prompts in LLM-assisted SE tasks and the challenges they face. As a pre-study, the results provide an initial empirical basis for the more extensive investigations planned in our research agenda.

The survey targeted professionals with experience using LLMs in SE contexts. Developers, testers, data scientists, architects, managers, and researchers were eligible to participate. The survey contained 18 questions on demographic background, prompt practices, and challenges, informed by our experience and related work~\cite{chen2025promptware, liang2025prompts, tafreshipour2025prompting}. Questions included multiple-choice, single-choice, five-point Likert scales, and open-ended formats. The latter enabled participants to elaborate on aspects such as the contextual information included in prompts or preferred structuring approaches. The survey was piloted with one developer and two researchers, resulting in refinements to improve clarity and consistency.

Data were collected in May 2025 using the Unipark platform\footnote{https://www.unipark.com} and distributed through professional networks, LinkedIn, and direct contacts. We received 74 valid responses, with no indications of incomplete or unserious participation. 

Given the sample size and scope, the analysis focused on descriptive statistics, including frequencies, percentages, and cross-tabulations, complemented by subgroup comparisons  to explore potential variations. Open-ended responses were analysed using lightweight thematic coding to identify recurring themes. These qualitative insights enrich the quantitative findings and provide a richer picture of current practices and challenges in prompt use for SE tasks. All material and data are available in \url{https://doi.org/10.5281/zenodo.15464269}.

\section{Preliminary Results}
\label{sec:results}

We collected 74 valid responses from professionals with experience using LLMs in SE contexts, mainly located in Germany (51\%), Brazil (26\%), and Sweden (12\%). Most respondents were developers (59\%), followed by researchers (31\%), and data scientists (22\%), with multiple roles allowed. The sample was mainly mid-career (55\% with 4–10 years’ experience) and 43\% practitioners.

\subsection{Prompt Practices}

Our results show that software professionals use LLMs frequently, with over 90\% of participants reporting daily or weekly use, primarily for code generation (77\%), bug fixing (55\%), writing documentation (50\%), and code explanation (49\%). Professionals also reported using prompts to find library functionality, compare approaches, and explore tools, essentially using LLMs as an alternative to search engines. Other reported activities include generating prompts themselves and brainstorming both technical and conceptual solutions to problems.

\textit{Prompt structure} typically combines elements such as task descriptions (93\%), code snippets (57\%), and project context (53\%). Participants also reported including constraints and examples, most of the time in a single detailed instruction (57\%). The amount of context typically added is moderate (50\%). Open-ended responses suggest that the level of context provided often depends on the task: for simple functions, a basic prompt may suffice, while test generations typically require more extensive input.

\textit{Prompt reuse} also remains uncommon: only 11\% reuse prompts regularly, while 46\% never do. The lack of modular structures renders it difficult to apply prompts across tasks, reinforcing a tendency to start from scratch with each new query. The use of \textit{prompt guidelines in SE contexts} appears rather mixed. While 34 participants reported following no guidelines, others referred to personal practices (26), online resources (20), or organization-wide standards (3). This variation suggests a lack of standardization and a high degree of individual experimentation. 

\textit{Prompt refinement} also emerged as a common part of the practice, with over 85\% of participants reporting that they refine prompts at least sometimes. Refinement is typically triggered by issues such as hallucinated output (69\%), unmet requirements (65\%), or vague responses (57\%). This behavior resembles a lightweight validation loop, where prompts are reworked until the output aligns with the user’s intent. Participants stopped refining prompts when the output is considered ``good-enough'' (68\%) or when it meets the requirements (58\%). While ``good-enough'' was not explicitly defined, this suggests that many practitioners rely on highly subjective, task-specific thresholds rather than formal correctness criteria. 

A closer look at participants with 10 or more years of experience (N=11) reveals some notable deviations from the overall trends. This group prefers using structured templates (6) and step-by-step instructions (4). Additionally, experienced professionals show greater reliance on external resources, with the majority (7) reporting the use of literature or internet-based guidelines. This suggests that more experienced professionals tend to adopt more structured and externally informed prompting approaches, possibly reflecting their preference for systematic and reusable practices.

\subsection{Prompt Challenges}

Participants reported a range of challenges when formulating prompts for SE tasks. We found that the most frequently cited difficulties include determining the appropriate level of detail (68\%) and ensuring that the LLM understands the provided context (47\%). These findings reflect the inherent ambiguity of prompt development: too little detail may lead to generic or incorrect outputs, while too much can overwhelm the LLM. Other common challenges included overcoming LLM limitations (42\%), translating requirements into clear instructions (36\%), and managing large requirements (30\%). Collectively, these challenges reveal a tension between the tacit knowledge embedded in SE and the explicit representations required to prompt LLMs effectively.

To navigate these difficulties, participants stated to employ a variety of strategies. The most common ones were making instructions more explicit (80\%), adding more contextual information (68\%), and breaking tasks into smaller sub-components (49\%). Additional strategies included giving examples, rephrasing prompts, and adding constraints on output style, highlighting a predominantly iterative approach to prompt refinement, often guided by trial-and-error.
\section{Conclusions and Discussion}

This work sets out a research agenda that positions prompts as SE artifacts. Our vision addresses three key aspects: (i) characterizing current prompt practices and influencing factors, (ii) analyzing prompts in terms of their evolution, traceability, and reuse, and (iii) developing evidence‑based guidelines for their management. As a first step, we conducted an exploratory survey with 74 software professionals from six countries. The findings reveal that while LLMs are already embedded in daily SE activities, prompting is highly ad‑hoc, shaped by individual experimentation rather than systematic practices. Reuse is rare, guidelines are inconsistently applied, and refinement is largely reactive, driven by issues such as hallucinated output and unmet requirements. These patterns point to a lack of structured support for sustaining prompt quality over time.

The survey also highlights challenges primarily linked to prompt engineering, such as determining the right level of detail, ensuring LLMs understand the provided context, and translating requirements into clear instructions. While these issues are closely tied to the creative and technical aspects of prompt formulation, they also have implications for prompt management since better documentation, reuse, and traceability could help mitigate them. 

These results provide empirical grounding for our central premise: that prompts should be treated as first‑class SE artifacts. Like other artifacts, they influence system behavior, evolve over time, and incur maintenance overheads. Managing them systematically has the potential to improve maintainability, reproducibility, and trust in LLM‑assisted development. However, our findings also make clear that any management approach must balance potential benefits with the perceived effort, addressing practitioners’ concerns about over‑engineering.



\bibliographystyle{splncs04}
\bibliography{references}

\end{document}